\RequirePackage{fix-cm}
\documentclass[smallextended]{svjour3}       
\smartqed  
\usepackage{amsmath,amsfonts,mathrsfs,subfig,graphicx}
\usepackage{nicefrac}
\usepackage{cancel}
\usepackage{overpic}
\usepackage[T1]{fontenc}

\graphicspath{{./pictures/}{./}}

\newcommand{\re}[1]{(\ref{eq:#1})}

\def\phi{\varphi}
\def\eps{\varepsilon}
\def\rho{\varrho}
\def\d{\mathrm{d}}
\def\p{\partial}

\renewcommand{\vec}[1]{\boldsymbol{#1}}

\newlength{\obrA} 
\newlength{\obrB} 

\newcommand{\Cm}{\mbox{C\,--\,metric}}

\newcommand{\sUP}[3]{\cancel{S}_{(#1,#2)}(#3)}
\newcommand{\sDN}[3]{\bcancel{S}_{(#1,#2)}(#3)}
\newcommand{\sDNs}[3]{\bcancel{S}^2_{(#1,#2)}(#3)}

\begin{document}
\title{On the nature of cosmic strings in black hole spacetimes}
\author{David Kofro\v{n}}
\institute{David Kofro\v{n} \at
              Institute of Theoretical Physics, Faculty of Mathematics and Physics,
Charles University, \\
V Hole\v{s}ovi\v{c}k\'{a}ch 2, 180\,00 Prague 8, Czech Republic
              \email{d.kofron@gmail.com}}

\date{Received: date / Accepted: date}
\maketitle
\abstract{
A new model for cosmic strings (i.e.\;conical singularities) attached to black holes is proposed. These string are obtained by a explicit construction via  limiting process from the so\,--\,called Bonnor rocket. This reveals quite surprising nature of their stress\,--\,energy tensor which contains first derivative of Dirac $\delta$ distribution. 
Starting from the Bonnor rocket we explicitly construct the Schwarzschild solution witch conical singularity and the \Cm. In the latter case we show that there is a momentum flux through the cosmic string, causing the acceleration of the black hole and the amount of this momentum is in agreement with the momentum taken away by gravitational radiation.
 }

\section{Introduction}
Among the many solutions of Einstein's equations with non-trivial topology, those possessing topological defects of various kind are of special interest. Beside their specific geometrical properties,
such spacetimes are often inspired by considerations in particle physics and cosmology. In particular, defects known as \emph{cosmic strings} typically arise as a result of spontaneous
symmetry breaking in Yang-Mills theories \cite{Nielsen-1973}. Phase transitions of this kind were conjectured to happen in the early universe \cite{Kibble-1976} and strings produced by this
mechanism could be sources of presently detectable gravitational radiation \cite{Vachaspati-1984,Vachaspati-1985}. Another intriguing feature of cosmic strings is the possibility
that their presence could be responsible for the actual existence of magnetic monopoles.

In this paper we do not intend to study the microscopic origin of strings and treat them from purely geometrical point of view. Topologically, the presence of a cosmic string is reflected in
the non-triviality of the first homotopy group of the spacetime, meaning that there exist closed curves surrounding the string that cannot be continuously deformed
to a point. Metrically, spacetimes exhibiting a cosmic string suffer from conical singularities in the planes transversal to the string \cite{Vilenkin-2000}. 

The cosmic string spacetimes --- flat vacuum spacetimes with conical singularity and \emph{cylindrical} symmetry ---  are known and for a long time. The structure of the singularity was first studied by Sokoloff and Starobinskii \cite{SokSta77} and the first attempt to model a cosmic string was due to Villenkin \cite{Vilenkin_1981} in the linearized theory. In the full general relativity Hiscock \cite{Hiscock_1985} reconstructed the same results. 

Hiscock used the Weyl form \cite{Weyl_1917} of static, cylindrically symmetric metric which is given by
\begin{equation}
\d s^2 = -e^{2\nu}\,\d t^2 + e^{2\lambda} \left( \d r^2+\d z^2 \right)+e^{2\psi} \d\phi^2\,,
\label{eq:}
\end{equation}
where $\nu$, $\lambda$ and $\psi$ are functions of $r$ only. 

He considered an infinite cylinder of matter with density $\rho=\eps$ and negative pressure $p=-\eps$ in the direction of the axis and vacuum outside of this cylinder.
Then the stress energy tensor reads
\begin{align}
T^t_t & = T^z_z = -\eps & \text{for}\ & r\leq l\,, \label{}\\
T^t_t & = T^z_z = 0 & \text{for}\ & r> l\,, \label{}
\end{align}
with all the other components trivially equal to zero. The constant $l$ defines the radius of the cylinder and the junction conditions by Israel \cite{IsraelSH,IsraelSHe} are imposed on the surface $r=l$. 

Then, in the limit when the radius of the cylinder is decreased and the density is increased simultaneously so that the ''mass per unit length'' (defined as integral of $T_{tt}$ over surfaces of constant $t,\,z$) remains constant a vacuum cosmic string spacetime is constructed, resulting in deficit angle around the axis.

In paper by Geroch and Trashen \cite{getrstrings} which deals with distributional sources in general relativity is shown that the sources of the cosmic strings cannot be found unambiguously (in the sense that there is no relation in between the deficit angle and the ``mass per unit length''). Let us present here their example briefly. They consider the same model as Hiscock, this time in cylindrical coordinates 
\begin{equation}
\d s^2 = -\d t^2 + \d z^2 + \d r^2 + \beta^2 (r)\, \d\phi^2\,,
\label{eq:GTm}\end{equation}
where 
\begin{equation}
\beta (r) = \begin{cases}
  \frac{l}{\gamma} \, \sin \left( \frac{\gamma r}{l} \right), & r\leq l\,,\rule[-1em]{0em}{1em} \\
  \left[ r-l+\frac{l}{\gamma}\, \tan\gamma \right]\cos\gamma\,, & r>l \,.
  \end{cases}
\label{eq:GT}\end{equation}
This metric is $C^1$ across $r=l$ (the surface of the cylinder) and $\gamma\in(0,\,2\pi)$ is a constant.
The stress energy tensor corresponding to the metric \re{GT} is identically zero for $r>l$ (vacuum) and for $r\leq l$ 
\begin{equation}
\boldsymbol{T}=\frac{\beta''(r)}{\beta(r)}\left[ \d t^2 - \d z^2 \right]=  -\frac{\gamma^2}{l^2}\left[ \d t^2 - \d z^2 \right]
\label{eq:}\end{equation}
which allows us to calculate the mass per unit length of the cylinder 
\begin{equation}
\mu_l = 2\pi \left(1-\cos\gamma\right).
\label{eq:mpul}
\end{equation}
The string (with a deficit angle given by $2\pi \gamma$) is obtained by taking the limit $l\rightarrow 0$.

But then they suggest the following modification of the metric \re{GTm} 
\begin{equation}
\d s^2 = e^{2\lambda f\left( \frac{r}{l} \right)} \left( -\d t^2 + \d z^2 + \d r^2 + \beta^2 (r)\, \d\phi^2 \right),
\label{eq:GTmod}\end{equation}
where $f$ is an arbitrary smooth nonnegative function with support on $\langle \nicefrac{1}{2},\,1\rangle$ (this modifies the matter content of the outer half of the cylinder, keeping the axis regular). The calculations of mass per unit length yield the following value
\begin{equation}
\mu_l = 2\pi\left( 1-\cos\gamma \right) - 2\pi \lambda^2 \int_0^1 \frac{\sin \gamma\, x}{x}\left[\, f'(x) \right]^2 \d x\,,
\label{eq:}\end{equation}
which is strictly less (but otherwise quite arbitrary up to the restrictions imposed on the function $f(x)$ above) than the value \re{mpul}, even though the strong energy condition is fulfilled (although the stress\,--\,energy tensor is more complicated).

Aside from these ambiguities more sophisticated matter models of cosmic strings have been proposed since then, see \cite{Christensen_1999,Dyer_1995,Sen_1997}, but all of them consider \emph{cylindrical} symmetry only.

Now, the same property --- deficit angle and, thus, the cosmic string --- is inevitably found also in the \Cm\ spacetime and can be implemented in an arbitrary \emph{axisymmetric} solution.

The question we raise in this paper is whether it is possible to interpret these strings (piercing the event horizon and extending to infinity) on the same level as a priori cylindrically symmetric cosmic strings? That is usually done by attributing ''mass per unit length'' to them.

To our best knowledge the only attempt to resolve string like sources in GR in general has been done by Israel \cite{Israel77}. His pioneering work suggests to investigate the strings in the near string limit. In this paper Israel himself considered, amongst other examples, static spacetime containing two black holes endowed with cosmic string which keeps them apart in Weyl coordinates. But his near axis limit somehow pushes aside the black hole horizon (which itself is degenerate in Weyl coordinates). 

In order to understand the origin of strings in the Schwarzschild solution or the \Cm\ one should provide a constructive and well-controlled procedure. We will do so in the following text. We do not attempt to provide a general treatment of string-like sources.

The Section \ref{sec:math} introduces relevant mathematical definitions employed later in the text and fixes the notation.

In Section \ref{sec:Bonnor} the starting point of our calculations --- the so called Bonnor rocket solution of Einstein field equations --- is reviewed. The Bonnor rocket \cite{Bonnor_1996} is a quite general black hole solution which contains an arbitrary axisymmetric and time dependent null dust along outgoing geodesic (being thus a generalization of Vaidya solution \cite{Vaidya}). The Schwarzschild solution, resp. the \Cm, as a particular example of the Bonnor rocket is treated in Subsection \ref{subsec:Schw}, resp. \ref{subsec:Cm}. Dynamical processes which lead to the string formation or to the smooth transition of Schwarzschild black hole to the \Cm\ are discussed.

The dynamical situations are difficult to treat, thus, the following Section \ref{sec:st} contains the detailed calculations of the structure of the string in a sequence of static spacetimes.

\section{Mathematical prerequisites} \label{sec:math}
\subsection{Step functions}
We shall use step functions of different profiles. One of them is the nonanalytic, yet smooth step function $S(x)$
\begin{align}
S(x) &=
\begin{cases}
0 & \text{for\ } x<x_0 \,, \\
\frac{f\left( \frac{x-x_0}{x_1-x_0} \right)}{f\left( \frac{x-x_0}{x_1-x_0} \right)+f\left(  1 - \frac{x-x_0}{x_1-x_0}\right)} &\text{for\ } x\in\langle x_0,\,x_1\rangle \,, \\
1 & \text{for\ } x>x_1 \,,
\end{cases}
\label{}
\end{align}
where $f(x) = e^{-\nicefrac{1}{x}}$, which is a step in between $x_0$ and $x_1$.

The another class of step functions, so called smooth-step functions, are polynomials of order $2n+1$ with boundary conditions prescribed by $f(x_0)=0$, $f(x_1)=1$ and  $\d^j f(x_0) / \d x^j = \d^j f(x_1) / \d x^j = 0$  for $j=1,\,2,\dots n$ which are simple to construct, manage analytically and are smooth up to the order $n$.

In general we will have a step up function $\sUP{a}{b}{x}$ which vanishes for $x<a$, has a desired interpolation in between 0 and 1 for $x\in\langle a,\,b\rangle$ and is equal 1 for $x>b$. The step down is then simply $ \sDN{a}{b}{x} = 1-\sUP{a}{b}{x}$. 

Also the ``table'' function 
\begin{align}
T_{(a,b,c,d)}(x)\quad &=\quad \begin{cases}
\quad  0                &\text{for } x<a \,,\\
\quad  \sUP{a}{b}{x}  &\text{for } x\in\langle a,\,b\rangle \,,\\
\quad  1                &\text{for } x\in\langle b,\,c\rangle \,,\\
\quad  \sDN{c}{d}{x}  &\text{for } x\in\langle c,\,d\rangle \,,\\ 
\quad  0                &\text{for } x>d \,,\\
 \end{cases} 
\label{eq:Table}
\end{align}
will be of use.

\subsection{Fourier\,--\,Legendre series} \label{subsec:Legendre}
$L^2$ functions on the interval $\langle-1,1\rangle$ can be expanded in the basis of Legendre polynomials as
\begin{equation}
f(x) = \sum_{n=0}^{\infty} a_n P_n(x) \,,
\label{eq:}\end{equation}
where, due to the normalization of Legendre polynomials, the coefficients $a_n$ are
\begin{equation}
a_n = \frac{2n+1}{2}\, \int_{-1}^{1} f(x) P_n(x)\, \d x \,.
\label{eq:}
\end{equation}
The expansion of the Dirac $\delta$ distribution to this basis is given by (\cite{NIST} (1.17.22))
\begin{equation}
\delta (x-a) = \sum_{n=0}^{\infty} \left( n+\nicefrac{1}{2} \right) P_n(x) \, P_n(a) \,.
\label{eq:deltaFL}
\end{equation}

Legendre polynomials arise as the result of Gramm\,--\,Schmidt orthogonalization of monomials $\{x^j,\,j=0\ldots \infty\}$. In the following calculation the inverse relation is necessary
\begin{equation}
x^k = \sum_{j=0}^k a^{(k)}_j P_j(x) \,.
\label{eq:}\end{equation}
This will allow us to rewrite polynomial series in term of Legendre polynomials and sum them up.

The explicit formulae (which differ for even and odd powers of $x$) we found to be
\begin{equation}
\begin{split}
x^{2k} & = \sqrt{\pi}\; \frac{\Gamma(2k+1)}{2^{2k+1}} \;
\sum_{j=0}^{k} \tilde{a}^{(2k)}_{2j}\, P_{2j}(x) \,, \\
\tilde{a}^{(2k)}_{2j} & =             \frac{4j+1}{\Gamma(k+j+\nicefrac{3}{2})\, \Gamma(k-j+1)}\,, 
\end{split}
\label{eq:invLeg}
\end{equation}
for even powers of $x$ and
\begin{align}
x^{2k+1} &= \sqrt{\pi}\; \frac{\Gamma(2k+2)}{2^{2k+2}} \;
            \sum_{j=0}^{k} \tilde{a}^{(2k+1)}_{2j+1}\;P_{2j+1}(x)\,, \nonumber \\
\tilde{a}^{(2k+1)}_{2j+1} & = \frac{4j+3}{\Gamma(k+j+\nicefrac{5}{2})\, \Gamma(k-j+1)} \;,
\label{eq:}
\end{align}
for odd powers.
\vspace{1em}

\section{Bonnor rocket} \label{sec:Bonnor}
In 1996 Bonnor \cite{Bonnor_1996} found an explicit solution of Einstein field equations with null dust in which the central black hole can radiate the null dust with an arbitrary axisymmetric pattern and time profile. This solution belongs to the Robinson\,--\,Trautman class and thus posses an expanding null geodesic congruence which is shear free and twist free. The modern version of this metric can be found in \cite{GrPoB} and reads as follows
\begin{multline}
\d s^2 = - \left( -\frac{1}{2}\,G_{,xx}-\frac{2m(u)}{r}-r\left( bG \right)_{,x}-b^2Gr^2 \right)\d u^2 \\
 -2\,\d u\,\d r + 2br^2\,\d u\,\d x + r^2\left( \frac{\d x^2}{G}+G\,\d\phi^2 \right),
\label{eq:BR}
\end{multline}
where
\begin{align}
b(x,u) &= -A(u) -\int\frac{G_{,u}(x,u)}{G^2(x,u)}\,\d x\,, \label{eq:b}\\
G(x,u) &= \left( 1-x^2 \right)\left[ 1+\left( 1-x^2 \right)\tilde{h}(x,u) \right], 
\label{eq:origG}
\end{align}
with $A(u)$ an arbitrary function of $u$ and $\tilde{h}(x,u)>-1$ an arbitrary smooth bounded function.
This represents a Bonnor rocket, a particle emitting null dust (pure radiation) with the angular (as the $x=\cos\theta$) dependence
\begin{equation}
4\pi\, n^2(x,u) = -\frac{1}{8}\,\left( GG_{,xxx} \right)_{,x}+\frac{3}{2}\,m\left( bG \right)_{,x}-m_{,u} \,.
\label{eq:n}
\end{equation}
Corresponding stress energy tensor reads
\begin{align}
T_{ab} &= \rho\; l_a l_b  \,, & 
\rho   &= \frac{n^2}{r^2} \,, &
l^a    &= \left(\frac{\p}{\p r}\right)^a.
\label{eq:}
\end{align}

The form of $G(x,\,u)$ given by \re{origG} is not the most general one. It had been chosen by Bonnor so that the axis is regular. And, clearly, it does not guarantee that the quantity $n^2$ defined in \re{n} is positive.

Let us relax these restrictions, first of all we will consider the function 
\begin{equation}
G(x,\,u) = \left( 1-x^2 \right)\left( 1+ h(x,\,u) \right)
\label{eq:}
\end{equation}
and then we will omit the second power in the definition of $n(x)$.

Investigating the regularity condition \cite{Stephani2003} of the axis ($x=\pm 1$) given in terms of the norm of the axial Killing vector $\vec{\xi}_{(\phi)}$
\begin{equation}
\frac{1}{4} \lim_{x\rightarrow \pm 1} \frac{F_{,a}F^{,a}}{F} = 1+h(\pm 1,\,u)\,,
\label{eq:}
\end{equation}
where $ F=\vec{\xi_{(\phi)}}\cdot\vec{\xi_{(\phi)}}$, it is clear the function $h(x,\,u)$ determines the regularity of the axis.

The Bonnor rocket \re{BR}\,--\,\re{origG} was fine tuned and thus in its original form does not contain conical singularities. We can introduce them by rescaling $G\rightarrow KG$ which is equivalent to the choice $h(x,\,u)=\,$const. 
 
Then, scaling the coordinates and parameters as 
\begin{align}
\tilde{u}& = \sqrt{K}\,u \,, &
\tilde{r}& = r/\sqrt{K} \,, \\
\tilde{m}& = m/K\sqrt{K} \,, &
\tilde{A}& = A/\sqrt{K}\,,
\end{align}
leads to the same form \re{BR} the metric, except the term $K^2\d\phi^2$ which shows the presence of conical singularity as we consider $\phi$ to run form $0$ to $2\pi$ strictly.

This ``relaxed'' class of Bonnor rockets contains as a special cases Schwarzschild solution with conical singularities and the \Cm.

\subsection{The Schwarzschild solution} \label{subsec:Schw}
The metric \re{BR} with $b=0$ and $G=(1-x^2)$ is the Schwarzschild solution. For $b=0$ and $G=K(1-x^2)$ and after the aforementioned rescaling the Schwarzschild black hole with the horizon pierced by cosmic string is obtained. (For regular Schwarzschild solution $x=\cos\theta$ where $\theta$ is standard polar coordinate on sphere.) 

\begin{figure*}
\begin{center}
\subfloat[Schwarzschild]{
\begin{overpic}[width=1.0\obrB,keepaspectratio]{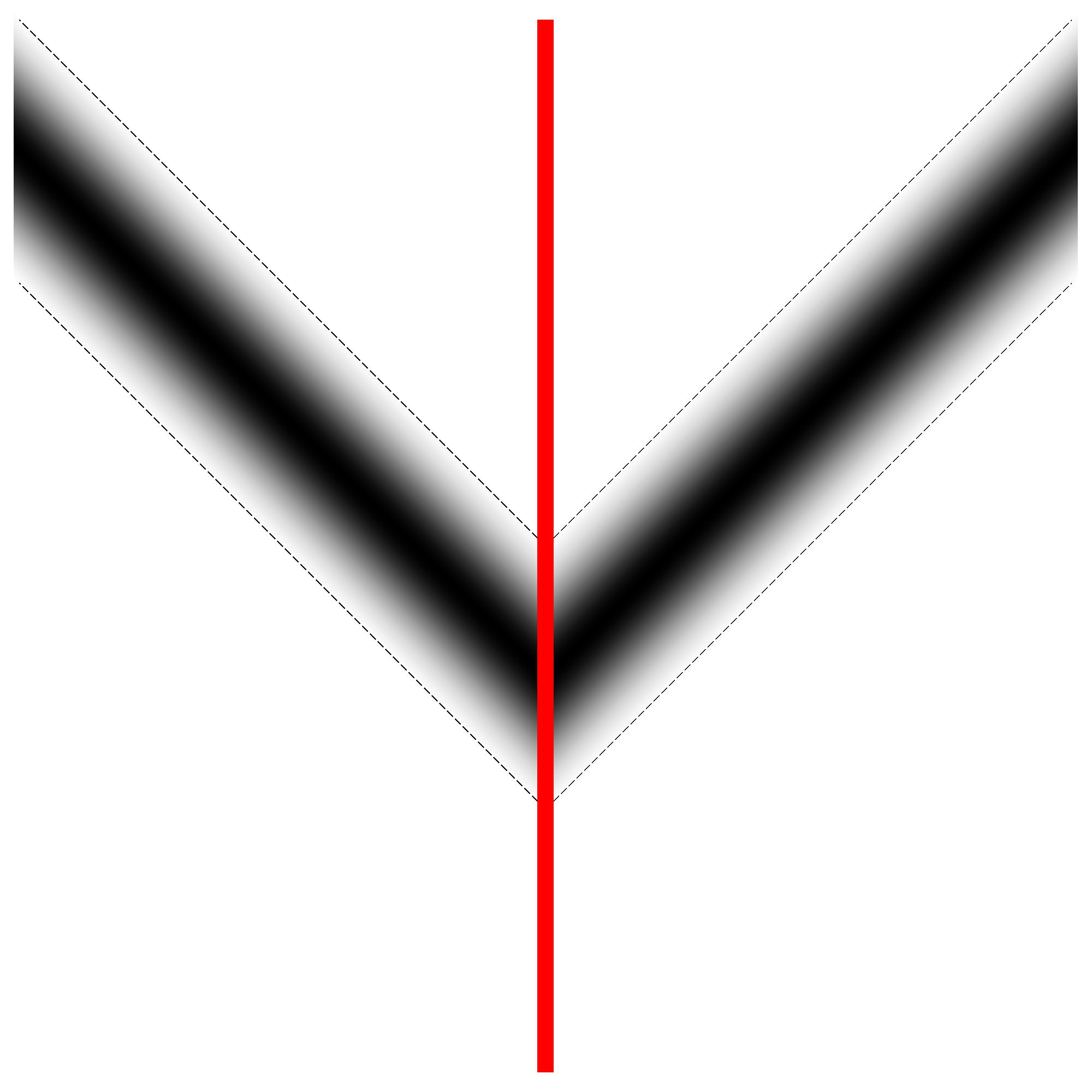}
\put(20,77){\rotatebox[origin=c]{-45}{$u=u_1$}}
\put(10,50){\rotatebox[origin=c]{-45}{$u=u_0$}}
\end{overpic}
} 
\qquad \qquad \qquad
\subfloat[Schwarzschild $\longrightarrow$ \Cm]{
\begin{overpic}[width=1.0\obrB,keepaspectratio]{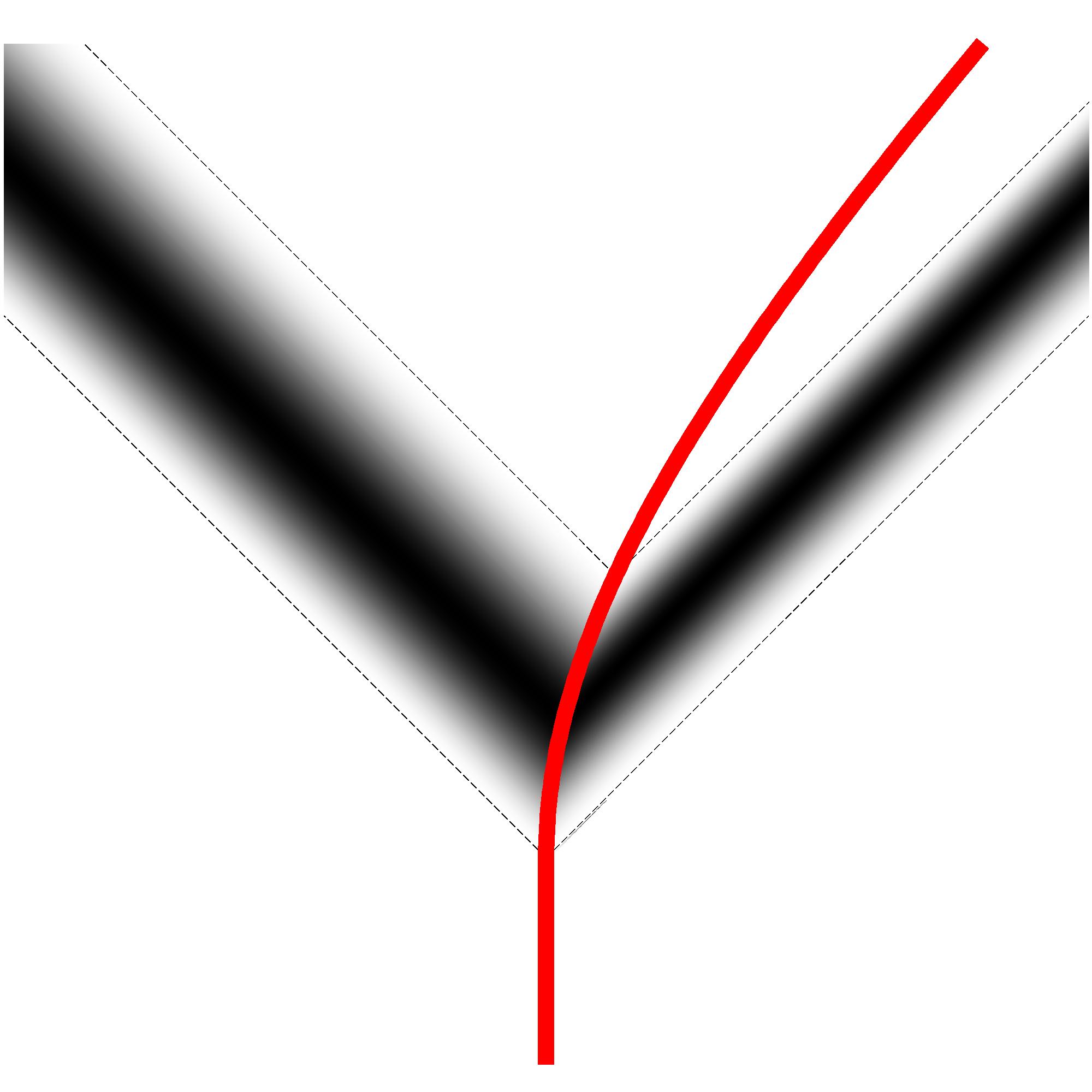}
\put(32,70){\rotatebox[origin=c]{-45}{$u=u_1$}}
\put(11,44){\rotatebox[origin=c]{-45}{$u=u_0$}}
\end{overpic}
}
\end{center}
\caption{Space-time diagram of dynamical formation of the cosmic string in the Schwarzschild solution, Fig. (a), and the smooth transition (nonuniform acceleration) of  the Schwarzschild black hole to the \Cm, Fig. (b), accompanied with the formation of cosmic strings with different stress energy tensors along the north and south poles.\\ 
The black hole horizon is depicted by bold red line. At $u=u_0$ the radiation phase starts, the gray scale represents the time evolution of the intensity of the radiation, and at $u=u_1$ the spacetime becomes static again, with conical singularities present.}
\label{fig:acc}
\end{figure*}

Using
\begin{align}
G(x,\,u) &= \left( 1-x^2 \right)\left( 1+ 2w \sUP{u_0}{u_1}{u} e^{-\frac{\sDN{u_0}{u_1}{u}}{1-x^2}} \right), \nonumber \\
A(u) &=0\,.
\label{eq:GSchw}
\end{align}
we get a transition between Schwarzschild for ($u<u_0$) through a radiating phase $u\in\langle u_0,\, u_1\rangle$ during which the axis is still regular, to a Schwarzschild pierced by cosmic string with $K=1+2w$ which appears at $u=u_1$ and there is no evolution later, see Figure \ref{fig:acc} (a) for a  schematic picture.

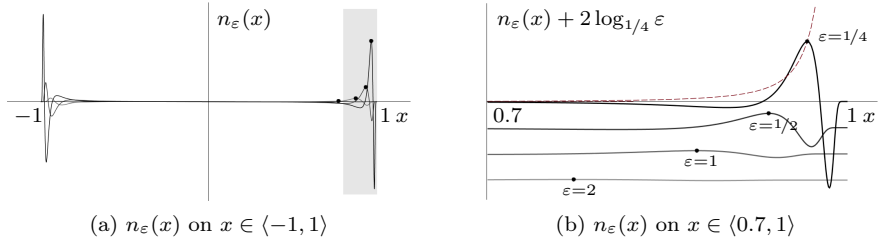
\begin{figure*}[t]
\centering
\subfloat[$n_\eps(x)$ on $x\in\langle -1,1\rangle$]{
\begin{overpic}[width=1.5\obrB,keepaspectratio]{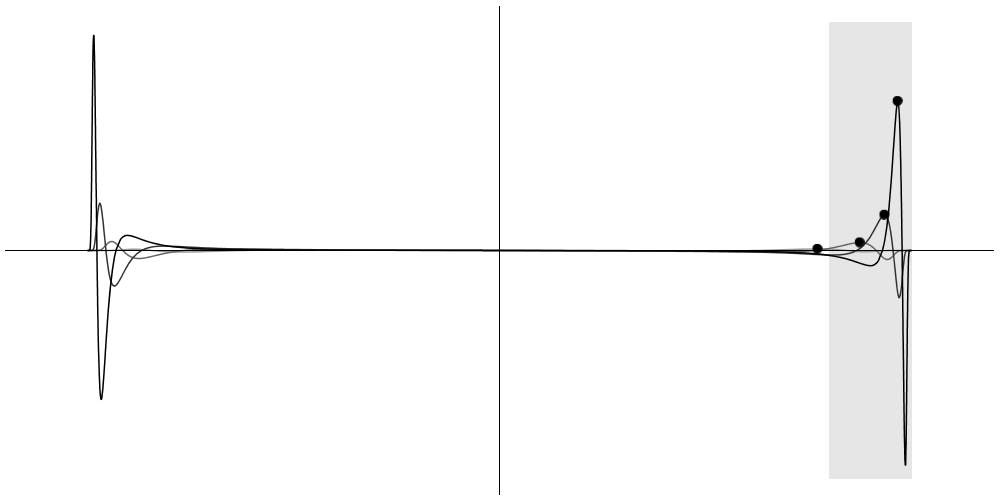}
\put(52,44){$n_\eps (x)$}
\put(96,20){$x$}
\put(2,20){$-1$}
\put(92,20){$1$}
\end{overpic}
} \qquad 
\subfloat[$n_\eps(x)$ on $x\in\langle 0.7,1\rangle$]{
\begin{overpic}[width=1.5\obrB,keepaspectratio]{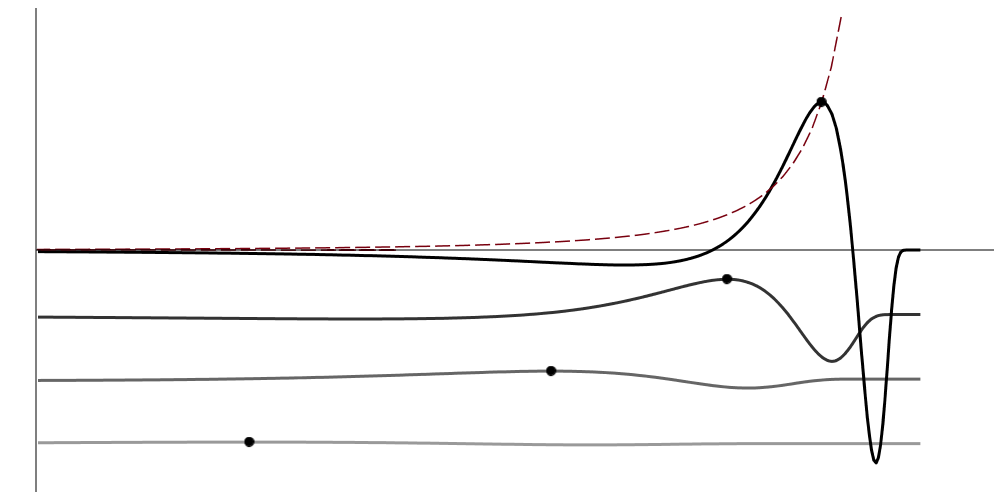}
\put(6,44){$n_\eps(x) + 2\log_{\nicefrac{1}{4}}\varepsilon$}
\put(96,20){$x$}
\put(5,20){$0.7$}
\put(92,20){$1$}
\put(22,2){$\scriptstyle{\eps=2}$}
\put(52,9){$\scriptstyle{\eps=1}$}
\put(68,18){$\scriptstyle{\eps=\nicefrac{1}{2}}$}
\put(85,40){$\scriptstyle{\eps=\nicefrac{1}{4}}$}
\end{overpic}
} 
\caption{An example of radiation pattern $n_\eps(x)$ and its focusing properties. These particular profiles are calculated for the structure function $G(x)$ given by \re{expCm} for $\eps=(2,1,\nicefrac{1}{2},\nicefrac{1}{4})$. In figure $(a)$ the plot for the whole angle is shown, whereas in $(b)$ the grey patch of $(a)$ is zoomed. Also, for the sake of clarity, there is an offset in the $y$ axis for every value of $\varepsilon$. The maxima of a every curve is depicted and the enveloping curve of these maxima is shown by dashed line, which is continuous function of $\varepsilon$ (with zero offset in the $y$ axis; thus it in this picture passes just through the maxima of the curve for $\varepsilon=\nicefrac{1}{4}$).}
\label{fig:rp}
\end{figure*}

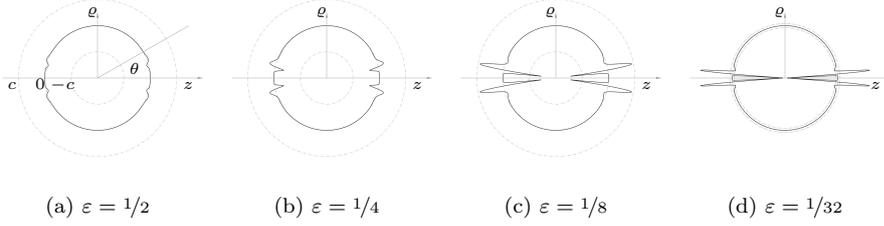
\begin{figure*}
\begin{center}
\subfloat[$\eps=\nicefrac{1}{2}$]{
\begin{overpic}[width=.79\obrB,keepaspectratio]{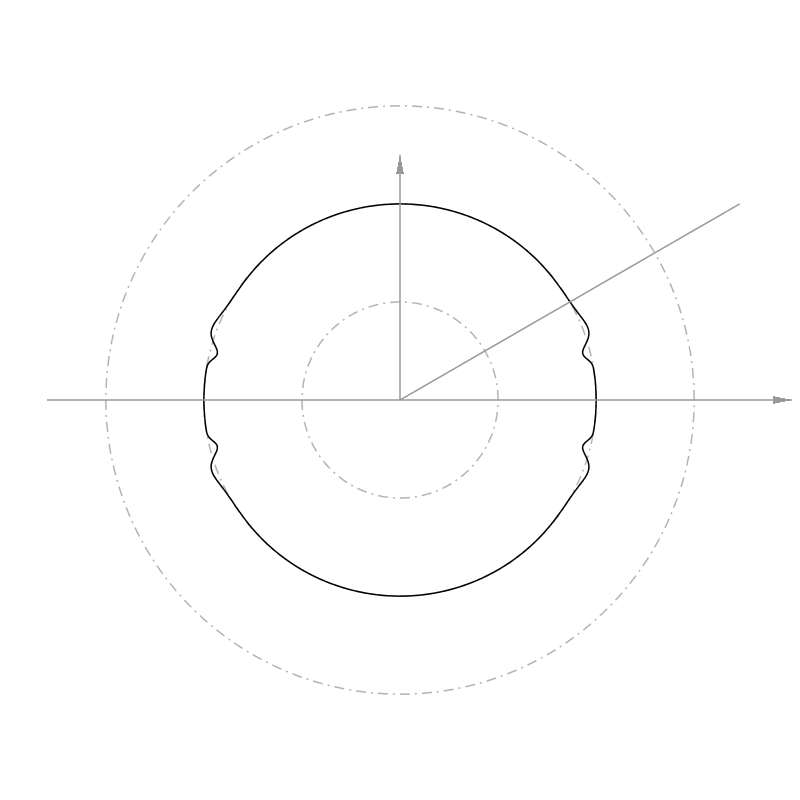}
\put(90,45){$\scriptstyle z$}
\put(65,52){$\scriptstyle \theta$}
\put(45,80){$\scriptstyle \rho$}
\put(8,45){$\scriptstyle c$}
\put(21,45){$\scriptstyle 0$}
\put(28,45){$\scriptstyle -c$}
\end{overpic}
} 
\subfloat[$\eps=\nicefrac{1}{4}$]{
\begin{overpic}[width=.79\obrB,keepaspectratio]{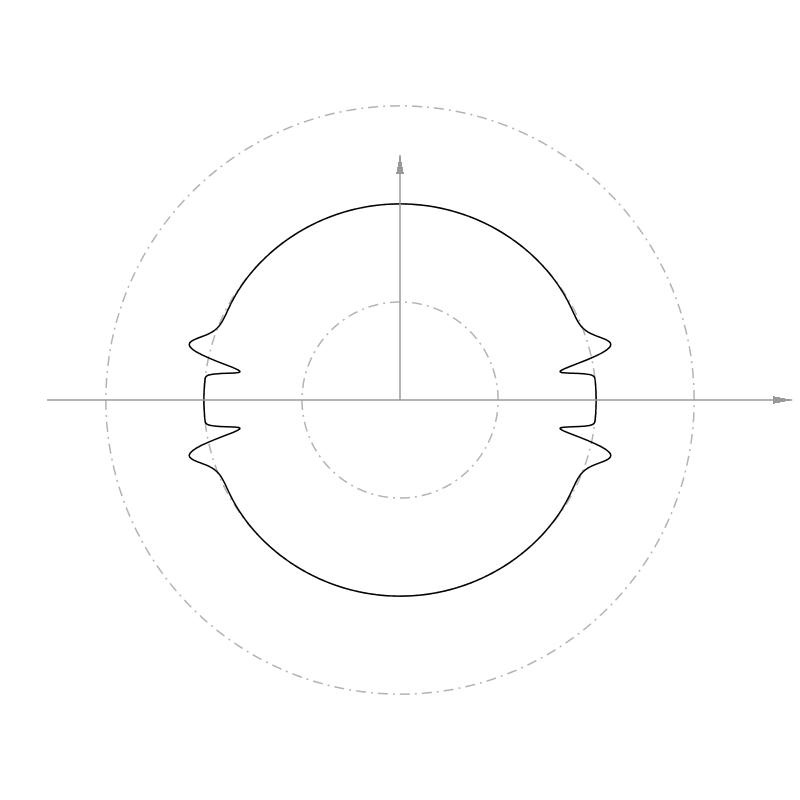}
\put(90,45){$\scriptstyle z$}
\put(45,80){$\scriptstyle \rho$}
\end{overpic}
}
\subfloat[$\eps=\nicefrac{1}{8}$]{
\begin{overpic}[width=.79\obrB,keepaspectratio]{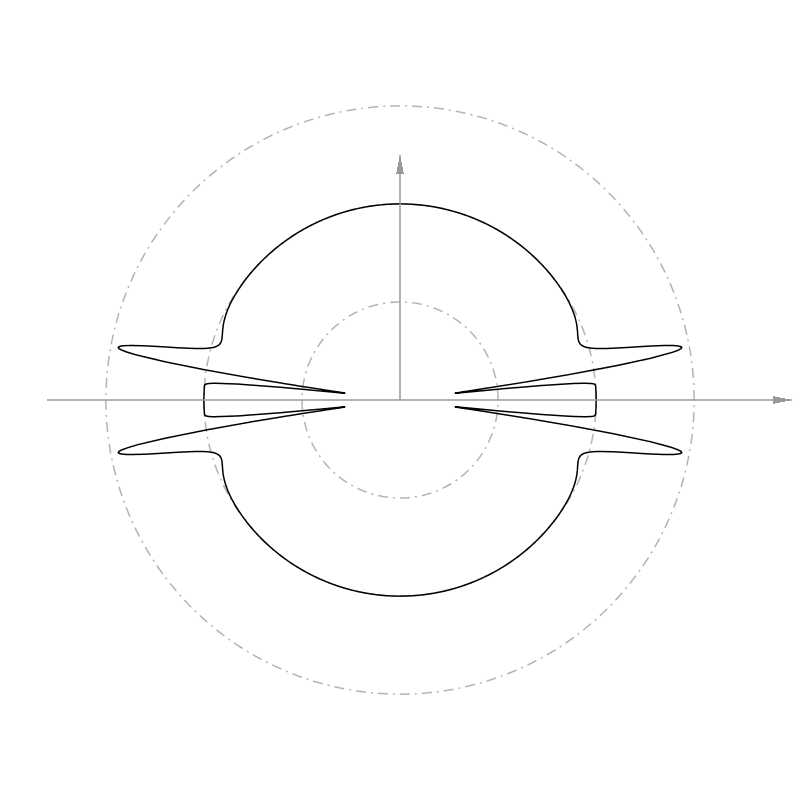}
\put(90,45){$\scriptstyle z$}
\put(45,80){$\scriptstyle \rho$}
\end{overpic}
}
\subfloat[$\eps=\nicefrac{1}{32}$]{
\begin{overpic}[width=.79\obrB,keepaspectratio]{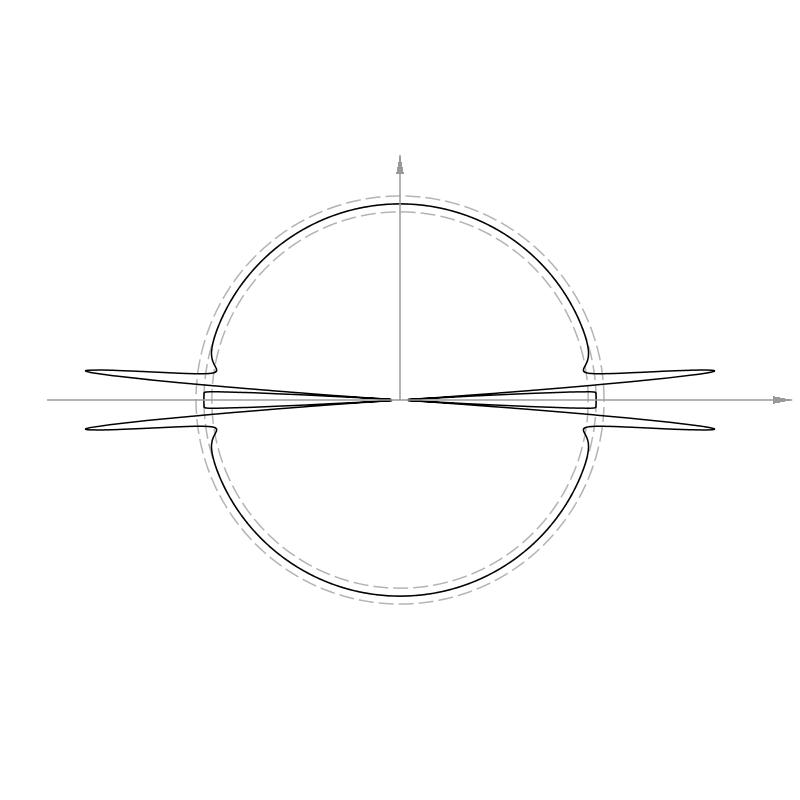}
\put(90,45){$\scriptstyle z$}
\put(45,80){$\scriptstyle \rho$}
\end{overpic}
}
\end{center}
\caption{The radiating pattern $n_\eps(x=\cos \theta)$ for the structure function \re{GSchwE} with $w=-\nicefrac{1}{10}$ depicted in polar plot. As the function $n_\varepsilon(x)$ is not strictly positive for all $x\in\langle -1,1\rangle$ an offset has been introduced as is clear from Fig. $(a)$ --- the value $c$ is at the outermost dash-dotted circle, zero is represented by the middle circle and $-c$ is represented as the innermost circle (the origin of coordinates is in the centre). In this case the value of $c=-400$. The focusation of radiation along the $z$ axis is clearly visible. In Fig. $(d)$ the scaling had to be readjusted so the circles are closer to each other. The visualisation of $n(x,u)$, Eq.~\re{n}, with $G(x,u)$ as in \re{GSchw} is visually indistinguishable.}
\label{fig:radpol}
\end{figure*}

Investigating the radiation pattern \re{n} we can see that the radiation gets more and more focused (see Fig. \ref{fig:rp} for an example for the \Cm or the Fig. \ref{fig:radpol} for the Schwarzschil solution -- the latter one is in polar coordinates and thus some of the properties are better readable from the picture) until the string appears and propagates to the infinity along a null world-line. 

\subsection{The \Cm} \label{subsec:Cm}
The \Cm\ in Robinson\,--\,Trautman coordinates \cite{GrPoB} (Eq. 19.4 therein) reads 
\begin{multline}
\d s^2 = -2\,\d u\,\d r +A^2r^2G\left( x-\frac{1}{Ar} \right) \d u^2\\
 - 2Ar^2\, \d u\, \d x + r^2\left( \frac{\d x^2}{G(x)}+G(x)\,\d\phi^2 \right),
\label{eq:cm}
\end{multline}
with
\begin{equation}
G(x) = \left(1-x^2\right)\left( 1+2Amx \right).
\label{eq:}
\end{equation}

Clearly, the metric element \re{BR} of the Bonnor rocket contains the \Cm\ \re{cm} as a special case, we simply have to set
\begin{align}
m(u) &=m\,, \\ 
b(x,u) &= -A\,, \\
G(x,u) &= \left( 1-x^2 \right)\left( 1+2Amx \right).
\label{}
\end{align}
Choosing the functions $G(x,\,u)$ and $A(u)$ in general Bonnor rocket metric \re{BR} as
\begin{align}
G(x,\,u) &= \left( 1-x^2 \right)\left( 1+2A(u)m x \,e^{-\frac{\sDN{u_0}{u_1}{u}}{1-x^2}} \right) \,, \nonumber \\
A(u) &= A\sUP{u_0}{u_1}{u} \,,
\label{eq:Gg}
\end{align}
we get a smooth transition from the static Schwarzschild solution for $u<u_0$, through a dynamic radiation phase for $u\in\langle u_0,\,u_1\rangle$ during which the radiation gets more a more focused along the still regular axis (but this time this radiation pattern is not reflection symmetric) to the \Cm\ for $u>u_1$. See Figure \ref{fig:acc} (b) for schematic picture. The axis start to posses a conical singularity at $u=u_1$ when the radiation is completely focused into an infinitely narrow beam and this singularity propagates along the null direction to infinity.

This dynamically obtained \Cm\ is for $u>u_1$ diffeomorphic to the \Cm\ but, clearly, cannot be analytically extended and does not contain the second black hole accelerated in the opposite direction.

The term 
\begin{equation}
\int \frac{G(x,u)_{,u}}{G^2(x,u)}\, \d x
\end{equation} hidden in the definition of the function $b(x,\,u)$ and thus in the radiation pattern $n(x,\,u)$, see \re{n}, is difficult, even impossible, to threat analytically. Therefore, in the next section, we will investigate these spacetimes as a sequence of different static spacetimes parameterized either by continuous parameter $\eps$ or integer $N$.

\begin{figure}
\centering
\begin{overpic}[width=0.8\obrA,keepaspectratio]{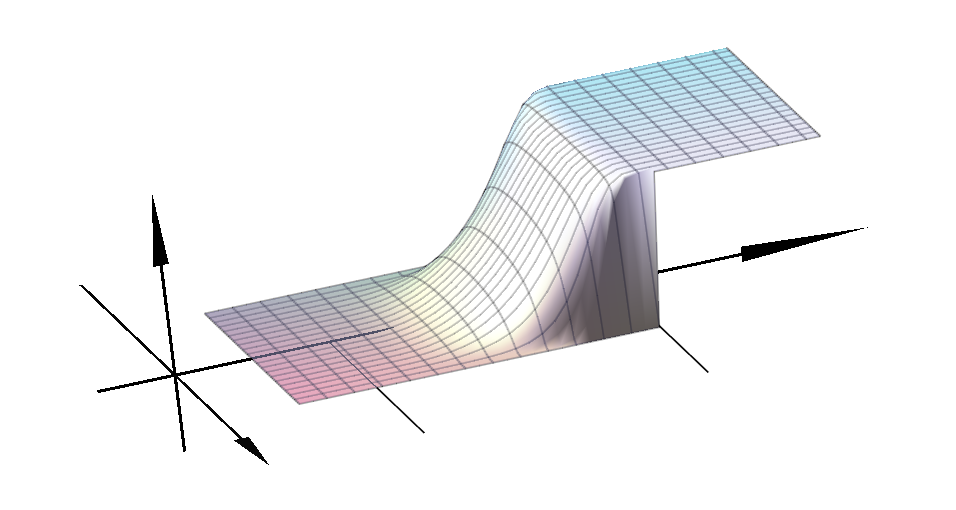}
\put(8,35){\rotatebox[origin=c]{10}{$h$}}
\put(20,3){\rotatebox[origin=c]{-45}{$x$}}
\put(85,23){\rotatebox[origin=c]{20}{$u$}}
\put(45,7){\rotatebox[origin=c]{20}{$u_0$}}
\put(68,13){\rotatebox[origin=c]{20}{$u_1$}}
\end{overpic}
\caption{The function $h=\sUP{u_0}{u_1}{u}\, e^{-\frac{\sDN{u_0}{u_1}{u}}{1-x^2}}$ which enters the structure function $G(x,\,u)$ and changes it dynamically in between $u_0$ and $u_1$.}
\label{fig:step}
\end{figure}
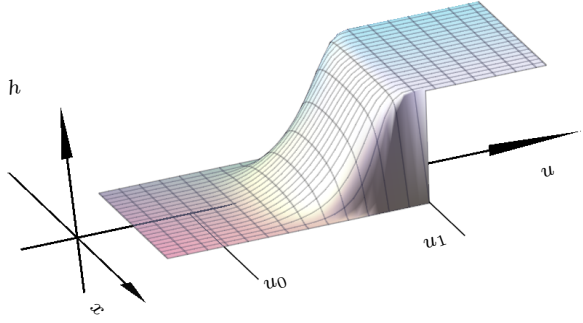

\section{Static treatment} \label{sec:st}
The Bonnor rocket emits null radiation and thus loses its mass, given by the energy outflow
\begin{equation}
-\frac{\d\, m(u)}{\d u} = \oint r^2 \rho\, \d\Omega = \oint n(x)\, \d\Omega \,.
\label{}
\end{equation}
Even if there is no time evolution there can be pure radiation, therefore our demand for staticity requires
\begin{align}
\oint n(x)\, \d \Omega &= 0\,, & m(u) &=m \,.
\label{eq:stac}
\end{align}
From this follows that $n(x)$ cannot be positive for $\forall x \in \langle -1,\,1\rangle$.

The arbitrariness in the choice of function $h(x)$ is almost infinite. Let us consider a sequence of spacetimes labelled by $N\in(\mathbb{N}\cup 0)$ given by
\begin{equation}
\label{eq:Schw-N}
\begin{split}
G(x,u) &= \left( 1-x^2 \right)\left( 1+2w\left( 1-x^{2N} \right) \right), \\
A(u) &=0 \,.
\end{split}
\end{equation}
which can be treated completely analytically.

For $N=0$ we get a standard Schwarzschild solution while in the limit $N\rightarrow \infty$ the Schwarzschild solution with a cosmic string is obtained. During the limiting process the axis is regular all the time.
 
Evaluating the radiation pattern \re{n} for the structure function \re{Schw-N} leads trivially to  
\begin{multline}
4\pi\, n_N(x) =        -2 w^2\, (4N+1)(2N+1)(N+1)N\, x^{4N}  \\
                      +4 w^2\, (2N^2+1)(4N-1)N\, x^{4N-2} \\
                       -2 w^2\, (4N-3)(2N-1)N(N-1)\, x^{4N-4} \\
                    +(2w+1)w\, (2N+1)^2\,(N+1)N\, x^{2N}  \\
                    -2(2w+1)w\, (2N^2+1)(2N-1)N\, x^{2N-2} \\
                    +(2w+1)w\, (2N-1)(2N-3)N(N-1)\, x^{2N-4}\,.
\label{eq:rpN}
\end{multline}

In this explicit and exact form the monomials $x^j$ can be expressed (or expanded) in the basis of Legendre polynomials as shown in the Section \ref{subsec:Legendre}. Then the limit $N\rightarrow \infty$ of $n_N(x)$ leads to  
\begin{equation}
4\pi\, n(x) = \left( w^2+w \right) \sum_{n=0}^{\infty} 2n\left( 2n+\nicefrac{1}{2} \right)\left( 2n+1 \right) P_{2n}(x) \,.
\label{eq:serSchw}
\end{equation}
This series can be summed up using the expansion of Dirac $\delta$ distribution \re{deltaFL} in the basis of Legendre polynomials from which we get
\begin{equation}
\Delta_+ \equiv \delta(x+1) + \delta(x-1) = 2\sum_{n=0}^{\infty} \left( 2n+\nicefrac{1}{2} \right)P_{2n}(x) \,.
\label{eq:}
\end{equation}
Now, employing the standard properties of Legendre polynomials and applying the following differential operator we get
\begin{multline}
\frac{1}{2}\, \frac{\d}{\d x}\left[ \left( 1-x^2 \right)\frac{\d}{\d x}\,\Delta_{+} \right]   =-\frac{\d}{\d x}\,\delta(x+1) +\frac{\d}{\d x}\,\delta(x-1)  \\
 =  -\sum_{n=0}^{\infty} 2n\left( 2n+\nicefrac{1}{2} \right)\left( 2n+1 \right) P_{2n}(x)\,,
\label{eq:DDp}
\end{multline}
in which we recognize the right hand side of \re{serSchw} and thus the final radiation pattern is
\begin{equation}
4\pi\,n(x) = -(w^2+w)\left[ \frac{\d}{\d x}\,\delta(x+1) -\frac{\d}{\d x}\,\delta(x-1) \right] .
\label{eq:DiracSchw}\end{equation}
This leads us to one of the main results of this paper -- to the explicit form of stress energy tensor for the cosmic string piercing the Schwarzschild black hole
\begin{equation}
{T}_{ab} = -(w^2+w)\left[ \frac{\d}{\d x}\,\delta(x+1) -\frac{\d}{\d x}\,\delta(x-1) \right] \frac{{l}_a {l}_b}{r^2}\,.
\end{equation}

Analogously, the \Cm\ can be obtained as a limiting case of the following sequence of spacetimes
\begin{equation}
\begin{split}
G(x,u) & = \left( 1-x^2 \right)\left( 1+2Amx\left( 1-x^{2N} \right) \right), \\
A(u) &= \left( 1-\frac{1}{N+1} \right) A,
\end{split}
\end{equation}
for which the condition \re{stac} of zero mass flux through an arbitrary sphere holds. Evaluating the radiation pattern is straightforward (but not short enough to be presented). Expressing monomials in the basis of Legendre polynomials and taking the limit $N\rightarrow \infty$ yields
\begin{multline}
4\pi\, n(x) =  A^2m^2 \sum_{n=0}^\infty 2n \left( 2n+\nicefrac{1}{2} \right)\left( 2n+1 \right) P_{2n}(x) \\
 + Am \sum_{n=0}^\infty \left( 2n+1 \right)\left( 2n+\nicefrac{3}{2} \right)\left( 2n+2 \right) P_{2n+1}(x) \,.
\label{eq:serCm}
\end{multline}
After some rearrangement of the expansion of the Dirac $\delta$ distribution in Legendre polynomials,
\begin{equation}
\Delta_- \equiv \delta(x+1) - \delta(x-1) = 2\sum_{n=0}^{\infty} \left( 2n+\nicefrac{3}{2} \right)P_{2n+1}(x) \,,
\label{eq:}
\end{equation}
and employing the properties of Legendre polynomials again we get
\begin{multline}
\frac{1}{2}\, \frac{\d}{\d x}\left[ \left( 1-x^2 \right)\frac{\d}{\d x}\,\Delta_{-} \right]  =-\frac{\d}{\d x}\,\delta(x+1) -\frac{\d}{\d x}\,\delta(x-1) \\
 =  -\sum_{n=0}^{\infty} \left( 2n+1 \right)\left( 2n+\nicefrac{3}{2} \right)\left( 2n+2 \right) P_{2n+1}(x)\,.
\label{eq:}
\end{multline}
As a result, we can recognize \re{serCm} to be 
\begin{multline}
4\pi\, n(x) =  -Am\left( Am+1 \right)\frac{\d}{\d x}\,\delta(x+1)  
  + Am\left( Am-1 \right)\frac{\d}{\d x}\,\delta(x-1) \,,
\label{eq:DiracCM}
\end{multline}
with the stress energy tensor given again by ${T}_{ab}=\rho\, {l}_{a}{l}_b$.

A different profile whose advantages lie in the fact that for $x\in\langle-1+\eps,\,1-\eps\rangle$ the spacetime is locally Schwarzschild or the \Cm\ can be found
\begin{equation}
G(x,u) = \left( 1-x^2 \right) \left( 1+2w \sUP{-1}{0}{-\eps}\; T_{(-1,-1+\eps,1-\eps,1)} (x) \right) \,,
\label{eq:Schw-e}
\end{equation}
for the Schwarzschild or
\begin{equation}
G(x,u) = \left( 1-x^2 \right)  \left( 1+2Am x \sUP{-1}{0}{-\eps}\; T_{(-1,-1+\eps,1-\eps,1)} (x) \right)  \,,\label{eq:Cm-e}
\end{equation}
for the \Cm. Step functions are now the polynomial smooth-step of order 7 or higher. In these cases we can calculate $n_\eps(x)$ and then, using computer algebra systems, its Fourier\,--\,Legendre expansion. In the next step --- in the limit $\eps \rightarrow 0^+$ we recover the results \re{serSchw} and \re{serCm}.

A completely different approach, which shows that these results are robust, is to use the functions
\begin{align}
G_\eps (x,u) &= \left( 1-x^2 \right)\left( 1+ 2w  e^{-\frac{\eps}{1-x^2}} \,\sDN{0}{1}{\eps} \right)\,, \nonumber\\
A(u) & = 0\,,
\label{eq:GSchwE}
\end{align}
for Schwarzschild
\begin{align}
G_\eps (x,u) &= \left( 1-x^2 \right)\left( 1+2Am x e^{-\frac{\eps}{1-x^2}}\,\sDN{0}{1}{\eps} \right)\,, \nonumber \\
 A(u) &= A\sDN{0}{1}{\eps}\,,
\label{eq:expCm}
\end{align}
for \Cm.

Evaluating the radiation pattern $n_\eps (x)$ is straightforward but it is impossible to express this function in the Fourier\,--\,Legendre series due to the integrals --- they consist of rational function multiplied by $e^{-\nicefrac{\eps}{(1-x^2)}}$.

For the Schwarzschild solution we get
\begin{equation}
4\pi\, n_\eps (x) = 
   - \frac{2w^2\sDNs{0}{1}{\eps}\;p_8(x)}{(x-1)^6(x+1)^6}\,\eps^2\,e^{-\frac{2\eps}{1-x^2}}  
    - \frac{w\sDN{0}{1}{\eps}\;q_8(x)}{(x-1)^6(x+1)^6}\,\eps^2\,e^{-\frac{\eps}{1-x^2}}\,, 
\label{}
\end{equation}
where $p_8(x)$ and $q_8(x)$ are polynomials
\begin{multline}
p_8(x) = 8\eps^2 x^4 - 2x^2\left( 1-x^2 \right)\left( 11x^2+9 \right)\eps \\
+3\left( 3x^4+8x^2+1 \right)\left( 1-x^2 \right)^2 ,
\end{multline}
and
\begin{align}
q_8(x) &= p_8(x)-4\eps^2 x^4 + 2x^2\left( 1-x^2 \right)\left( 7x^2+6 \right)\eps\,.
\label{}
\end{align}
Therefore, for now, consider the $n_\eps (x)$ as a distribution and let it act on test functions. We anticipate the result, of course. The behavior of $n_\eps (x)$ is governed by the term $e^{-\nicefrac{\eps}{(1-x^2)}}$, for $x\in (-1,\,1)$ the limit $\eps\rightarrow 0^+$ tends to 0.

In the radiation pattern $n_\eps (x)$ we can interpolate for small $\eps$ 
\begin{align}
e^{-\frac{\eps}{1-x^2}} & \sim e^{-\frac{1}{2}\frac{\eps}{x+1}}\,, &\text{for}\ & x\in\langle -1, 1);\ \text{denoted by}\ n^-_\eps\\
e^{-\frac{\eps}{1-x^2}} & \sim e^{\frac{1}{2}\frac{\eps}{x-1}}\,, &\text{for}\ & x\in (-1,\, 1\rangle;\ \text{denoted by}\ n^+_\eps \nonumber 
\label{}
\end{align}
and similarly for $e^{-\nicefrac{2\eps}{(1-x^2)}}$. 

Using computer algebra systems it can be analytically calculated how this distribution acts on basis of polynomials, i.e. evaluate the integral $n_\eps(x)\left[ x^N \right] = \int_{x_0}^1 x^N n_\eps(x)$. In the limit the result, independent on $x_0 \in (-1,\,1)$, is
\begin{align}
\lim_{\eps\rightarrow 0^+} \int_{x_0}^1 x^N n^+_\eps(x)\, \d x & = (w^2+w)N \\
& = -(w^2+w)\, \delta'(x+1) \left[ x^N \right]\,,\nonumber\\
\lim_{\eps\rightarrow 0^+} \int_{-1}^{x_0} x^N n^-_\eps(x)\, \d x & = (-1)^N N(w^2+w)  \\
&= -(w^2+w)\, \delta'(x-1)\left[ x^N \right]\,,\nonumber
\label{eq:}
\end{align}
and thus it acts as derivative of Dirac $\delta$ distribution as in \re{DiracSchw}.

The same procedure can be repeated for the \Cm\ with results as in \re{DiracCM}, of course.

So far we have calculated the stress energy tensor for the strings attached to the Schwarzschild black hole and the \Cm\ using three different regularization schemes with the same results. This shows that the procedure is robust.

Moreover, another interesting conclusion is at hand: in the case of the \Cm\ the null dust is not radiated away in a symmetric manner and thus carries the momentum away, in the rest frame of the black hole we find
\begin{equation}
P_z = \int_{-1}^1 n(x)\, x\; \d x = Am \,,
\label{eq:Pz}
\end{equation}
where $x$ is spherical harmonics $Y^0_1(x,\phi)$ and actually should be replaced by the solution of eigenfunctions on two sphere $t=\,$const and $r=\,$const as we have done in \cite{Kofron_2015,Kofron_2016}. Unfortunately the solution can be found only in terms of Heun general function and cannot be normalized. Yet, this solution for small $Am$ tend to $x$ and the corrections to $P_z$ given by \re{Pz} would be of order $A^2m^2$.

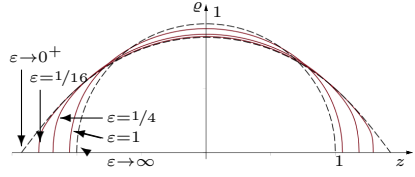
\begin{figure}
\centering
\begin{overpic}[width=1.5\obrB,keepaspectratio]{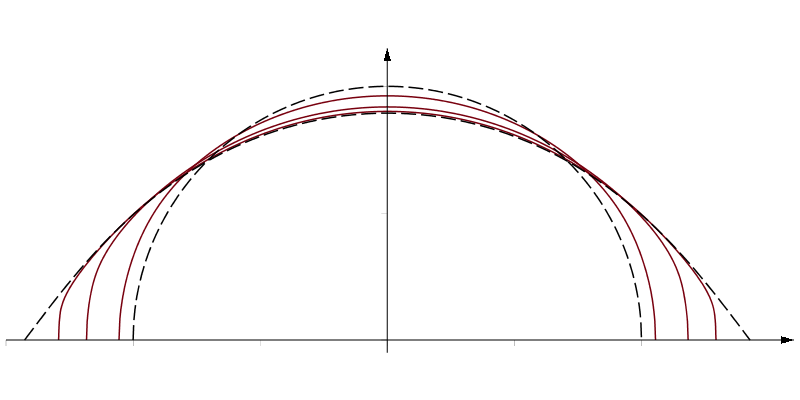}
\put(95,4){$\scriptstyle z$}
\put(80,4){$\scriptstyle 1$}
\put(24,15){$\scriptstyle \eps=\nicefrac{1}{4}$}
\put(23,16){\vector(-1,0){10.5}}
\put(24,10){$\scriptstyle \eps=1$}
\put(23,11){\vector(-4,1){7}}
\put(24,4){$\scriptstyle \eps\rightarrow\infty$}
\put(23,5){\vector(-3,2){5.5}}
\put(0,30){$\scriptstyle \eps\rightarrow 0^+$}
\put(3,28){\vector(0,-1){19}}
\put(5,25){$\scriptstyle \eps=\nicefrac{1}{16}$}
\put(7.6,23){\vector(0,-4){13}}
\put(45,43){$\scriptstyle \rho$}
\put(50,41){$\scriptstyle 1$}
\end{overpic}
\caption{The embedding of surface of constant $u$ and $r$ for the structure function \re{GSchwE} for various values of the parameter $\eps\in(0^+,\nicefrac{1}{16},\nicefrac{1}{4},1,\infty)$ and $w=-\nicefrac{4}{10}$. Basically the transition from a sphere ($\eps\rightarrow\infty$) --- a dashed halfcircle --- through a cigar shaped surfaces, $\varepsilon\in (1,\nicefrac{1}{4},\nicefrac{1}{16})$, with a regular axis --- in red --- into a sphere with cut-out angle ($\varepsilon\rightarrow 0^+$) and thus a singularity around poles is seen.}
\label{fig:emb}
\end{figure}

\vspace{1ex}

Finally, let us investigate the geometry of surfaces (with spherical topology) of constant $u$ and $r$ as embedded surfaces into $\mathbf{R}^3$. Assume its embedding in spherical coordinates $\d s^2= \d r^2 + r^2 (\d \theta^2 +r^2\d\phi^2)$ is of the form $r=R(\theta(x))$ (we apply the coordinate transformation $\theta=\theta(x)$ at the simultaneously). Then by comparison of the induced metrics 
\begin{align}
\d s^2 &= \left( R_{,x}^2+R^2\theta_{,x}^2 \right)\d x^2 + R^2\sin^2\theta\,\d\phi^2\,, \\
\d s^2 &= \frac{1}{G(x)}\,\d x^2+G(x)\,\d\phi^2\,,
\end{align}
we get the following differential equation for a newly introduced function $f(x)$
\begin{align}
R(x) &= \sqrt{f(x)^2+G(x)}\,, & f_{,x} &= \frac{\sqrt{-G\left( G_{,x}^2-4 \right)}}{2G}\,.
\label{}
\end{align}
In general the function $f(x)$ can be found in term of integral, the embedding in $\mathbf{R}^3$ is not possible if $G_{,x}^2>4$.

The embedding is the a surface of revolution defined by a curve in Euclidean coordinates 
\begin{equation}
[z,\,\rho]=\left[f(x),\,\sqrt{G(x)}\right]\,.
\end{equation}
 We numerically evaluate and plot these surfaces for $G(x)$ given by \re{GSchwE} in Fig. \ref{fig:emb}. And it can be nicely seen how the axis remains regular until the very last moment of the procedure.

\section{Israel's approach}
Let us briefly discuss the approach proposed by Israel \cite{Israel77} applied to our case. This approach relies on explicit construction of coordinates in the vicinity of the axis such that the metric is of the form
\begin{equation}
\d s^2 = \d \rho^2 + A^2(z,t)\d z^2 + B^2 \rho^2\d\phi^2 - C^2(z,t)\d t^2 \,,
\label{eq:}
\end{equation}
and investigating The extrinsic curvature $K_{ab}$ (and its densitized form $\mathscr{K}_a^b$) of cylinders of constant $\rho$
\begin{align}
K_{ab} & = \frac{1}{2}\,\frac{ \p g_{ab} }{\p\rho}\,, &
\mathscr{K}_a^b & = \sqrt{-\det g}\, K_a^b\,,
\label{eq:exc}
\end{align}
and its limit
\begin{equation}
\mathscr{C}_a^b = \lim_{\rho\rightarrow 0^+} \mathscr{K}_a^b \,.
\label{eq:}\end{equation}

Let us have a Schwarzschild solution endowed with cosmic string in Weyl coordinates
\begin{equation}
\d s^2 = -e^{2\psi}\, \d t^2 + e^{2\left(\lambda-\psi\right)}\, \left( \d r^2 + \d z^2 \right)+e^{-2\psi}\,r^2\,\d\phi^2 \,,
\label{eq:}\end{equation}
where
\begin{align}
\psi &= \frac{1}{2}\ln \left[ \frac{R_++R_--2m}{R_++R_-+2m} \right], &
\lambda &= \frac{1}{2}\ln \left[ \frac{\left( R_++R_- \right)^2-4m^2}{4R_+R_-} \right] + K \,,
\label{}
\end{align}
with $R_\pm = \sqrt{r^2+\left( z\pm m \right)^2}$. The parameter $K$ controls the regularity of the axis --- regularity condition reads $\lambda(0,z)=0$.

We can find the transformation from Weyl coordinates $(r,\,z)$ to approximate coordinates $(\rho,\,\zeta)$ in which $\rho$ is the affine parameter of geodesic connecting the axis with the point in its vicinity to an arbitrary order of precision (for $\zeta>m$) 
\begin{align}
r &= 0 
   + e^{-K}\sqrt{\frac{\zeta-m}{\zeta+m}}\, \rho 
   + \frac{e^{-3K}}{6}\frac{\sqrt{\zeta^2-m^2}m}{\left( \zeta+m \right)^4}\, \rho^3   \nonumber \\ 
&   - \frac{e^{-5K}}{120} \frac{\sqrt{\left( \zeta^2-m^2 \right)}m(2m+9\zeta)}{\left( \zeta+m \right)^7}\, \rho^5 + \dots \,,\\ 
z &= \zeta 
   - \frac{e^{-2K}}{2}\frac{m}{\left( \zeta+m \right)^2}\, \rho^2 
   - \frac{e^{-4K}}{24}\frac{m\left( m-3\zeta \right)}{\left( \zeta+m \right)^5} \rho^4 \nonumber \\
&    + \frac{e^{-6K}}{720}\frac{m\left( 35m^2+6m\zeta-45\zeta^2 \right)}{\left( \zeta+m \right)^8}\, \rho^6 + \dots \,,
\label{}
\end{align}
and then the Schwarzschild metrics reads
\begin{multline}
\d s^2  = -\frac{\zeta-m}{\zeta+m}\Biggl( 1 + me^{-2K}\,\frac{1}{\left(\zeta+m\right)^3}\,\rho^2 \\ 
\ -\frac{me^{-4K}}{12}\frac{ 13m-9\zeta }{\left( \zeta+m \right)^6}\,\rho^4 + o(\rho^6) \Biggr) \d t^2 \\
\  +\left( 1+ o(\rho^6)\right) \d \rho^2  + o(\rho^7)\, \d\rho\,\d\zeta \\
\  +\frac{\zeta+m}{\zeta-m}\Biggl( e^{2K} + \frac{m}{\left( \zeta+m \right)^3}\,\rho^2\\
\ +\frac{me^{-2K}}{12}\frac{7m-3\zeta}{\left( \zeta+m \right)^6}\,\rho^4 + o(\rho^6) \Biggr) \d\zeta^2 \\
  + e^{-2K}\rho^2 \left( 1-\frac{2}{3}\frac{e^{-2K}}{\left( \zeta+m \right)^3}\,\rho^2 + o(\rho^4) \right) \d\phi^2 \,.
\label{}
\end{multline}
The limit of densitized external curvature tensor is simply 
\begin{align}
\mathscr{C}_a^b&= \begin{pmatrix} 0 & 0 & 0 \\ 0 & 0 & 0 \\ 0 & 0 & 1 \end{pmatrix} \,,
\end{align}
in coordinates $(t,\,\zeta,\,\phi)$. But this tensor has null eigenvectors; and therefore the Israel's approach does not have to provide an decisive answer, as \emph{``Condition (vi) excludes ``lightlike'' sources which (like null surface layers) require special treatment''}\footnote{Citation from \cite{Israel77}.}. Although the axis itself is not a null hypersurface, the stress energy tensor is composed of null dust.

We have already seen in Section \ref{subsec:Schw} and \ref{subsec:Cm} that the nature of singularities is lightlike.

This shows that the conical defects are a very subtle subject which has to be treated carefully.

\section{Conclusions}
We have proposed a new model of cosmic strings attached to black holes and revealed their corresponding stress energy tensor in the case of Schwarzschild black hole and the \Cm. The strings are made of null dust.
 
Our explicit construction proved to be quite regularization independent (we used three different schemes).

For the \Cm\ the deficit angle is different on the north pole and on the south pole. This asymmetry suggests that there is a momentum flux through the cosmic strings. This flux has been calculated.

\begin{acknowledgement} 
D.K. acknowledges the support from the Czech Science Foundation, Grant 17-16260Y. Moreover, D.K. would like to thank Dr. M. Scholtz for inspiring discussions and comments on the manuscript.
\end{acknowledgement}

\bibliographystyle{spphys}
\bibliography{kofron}{}

\end{document}